\RequirePackage{fix-cm}
\documentclass[twocolumn,final]{svjour3}
\usepackage[utf8]{inputenc}
\usepackage[american]{babel}

\usepackage{graphicx}%
\usepackage{multirow}%
\usepackage{amsmath,amssymb,amsfonts}%
\usepackage{mathrsfs}%
\usepackage[title]{appendix}%
\usepackage[dvipsnames]{xcolor}
\usepackage{pifont}
\usepackage{textcomp}%
\usepackage{manyfoot}%
\usepackage{booktabs}%

\usepackage{threeparttable}
\usepackage{tabularx}
\newcolumntype{B}{X}
\newcolumntype{S}{>{\hsize=.5\hsize}X}
\newcolumntype{C}{>{\centering\arraybackslash}X}

\usepackage{array}
\usepackage{ragged2e}
\usepackage{xspace}

\usepackage{tikz}
\usetikzlibrary{arrows,shapes,patterns,decorations.text,decorations.pathreplacing}
\usetikzlibrary{automata,positioning}
\usetikzlibrary{trees}

\usepackage[frozencache,cachedir=_minted-output]{minted}
\BeforeBeginEnvironment{minted}{\medskip}
\AfterEndEnvironment{minted}{\medskip}
\usepackage{caption}
\usepackage{subcaption}

\newcommand{\LTL}[0]{\textsf{LTL}\xspace}

\newcommand{\MTL}[0]{\textsf{MTL}\xspace}
\newcommand{\STL}[0]{\textsf{STL}\xspace}

\newcommand{\PastLTL}[0]{\textsf{PastLTL}\xspace}
\newcommand{\PastLTLr}[0]{\textsf{PastRoLTL}\xspace}
\newcommand{\PastMTL}[0]{\textsf{PastMTL}\xspace}
\newcommand{\PastSTL}[0]{\textsf{PastSTL}\xspace}
\newcommand{\PastSTLr}[0]{\textsf{PastRoSTL}\xspace}
\newcommand{\PastFOLTL}[0]{\textsf{PastFOLTL}\xspace}
\newcommand{\PastFOMTL}[0]{\textsf{PastFOMTL}\xspace}
\newcommand{\PastFOSTL}[0]{\textsf{PastFOSTL}\xspace}

\newcommand{\textxt}[2]{{\scriptsize[\ensuremath{#1}:\ensuremath{#2}]}}

\begin{document}
\twocolumn[
\begin{@twocolumnfalse}
\title{Reelay: Online Temporal Logic Monitoring Framework}


\hyphenation{unbo-unded catego-rical da-ta}

\author{Dogan Ulus}


\institute{D. Ulus \at Bo\u gazi\c ci University, Istanbul, T\"urkiye\\\email{dogan.ulus@bogazici.edu.tr}}

\date{Received: date / Accepted: date}

\maketitle

\begin{abstract}
We present Reelay, a unified online temporal logic monitoring framework designed for the rigorous analysis and runtime verification of cyber-physical systems. Reelay addresses the fragmentation of existing logical formalisms and tools by providing a single computational model and interface that supports a broad class of temporal logics. These include Linear Temporal Logic (\LTL), Metric Temporal Logic (\MTL), and Signal Temporal Logic (\STL), along with their extensions for robustness semantics and first-order quantification over unbounded categorical data domains. At its core, Reelay translates temporal logic specifications into executable computation graphs operating as synchronous dataflow systems. This architecture ensures an efficient execution mechanism, making the framework ideal for high-frequency data streams regardless of behavior length. Uniquely, the framework supports both discrete and dense-time semantics, as well as delta-encoded temporal behaviors to minimize bandwidth usage in bandwidth-constrained environments. Reelay is implemented as a header-only C++ library with a high-level Python interface, facilitating integration across a wide range of deployment contexts, from resource-constrained embedded systems to autonomous robotic platforms. We demonstrate the practical applicability of the framework through a representative case study and performance experiments, illustrating how Reelay bridges the gap between expressive formal specifications and efficient runtime verification.


\keywords{Runtime verification \and Formal methods \and Temporal logic \and Real-time systems \and Cyber-physical systems}
\end{abstract}

\end{@twocolumnfalse}
]

\section{Introduction}
\label{sec:intro}

Temporal logic monitoring is a cornerstone of specification based runtime verification for complex real-time systems.
By continuously evaluating system traces against temporal logic specifications, these techniques ensure adherence to intended behaviors while detecting undesirable or unsafe outcomes in real time.
To meet the demands of modern cyber-physical systems, temporal logic monitoring frameworks must move beyond simple temporal ordering; they require high expressiveness to handle real-time constraints, complex event processing, and quantitative reasoning.

Linear-time Temporal Logic (\LTL)~\cite{pnueli1977temporal}, the pioneering application of temporal logic to software and system verification, provides a discrete-time formal framework for specifying the qualitative ordering of events.
While \LTL is sufficient for verifying many functional software properties, the rigorous analysis of real-time systems necessitates dense-time semantics, greater expressivity, and quantitative temporal reasoning.
Consequently, Metric Temporal Logic (\MTL)~\cite{koy90} emerged as a prominent extension, gaining widespread adoption in both academia and industry for specifying the time-critical behaviors of reactive systems.
Beyond these foundations, a diverse array of \LTL and \MTL extensions has been proposed to address specialized monitoring requirements. These advancements include Signal Temporal Logic (\STL) for real-valued signal predicates~\cite{maler2013monitoring}, robustness semantics for quantitative reasoning~\cite{donze2010robust}, and first-order quantification over unbounded categorical data domains~\cite{basin2018algorithms,havelund2020first}. However, despite these significant syntactic and semantic breakthroughs in the field~\cite{bartocci2018specification,sanchez2019survey}, the practical applicability of runtime verification remains limited by a highly fragmented landscape of logical formalisms and monitoring tools. This lack of unification creates a major barrier to industrial adoption, as practitioners are forced to navigate incompatible tools and conflicting semantics to verify modern, complex cyber-physical systems.

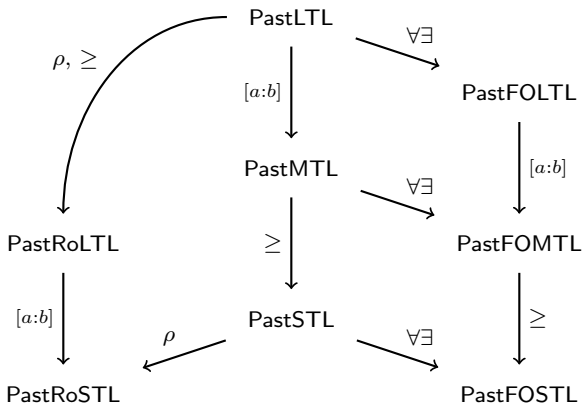
\begin{figure}[tb]
\begin{center}
\begin{tikzpicture}
\node (ltl) at (0,0) [inner sep=8pt]{\PastLTL};
\node (mtl) at (0,-2) [inner sep=8pt]{\PastMTL};
\node (stl) at (0,-4) [inner sep=8pt]{\PastSTL};
\node (foltl) at (3,-1) [inner sep=8pt]{\PastFOLTL};
\node (fomtl) at (3,-3) [inner sep=8pt]{\PastFOMTL};
\node (fostl) at (3,-5) [inner sep=8pt]{\PastFOSTL};
\node (ltlr) at (-3,-3) [inner sep=8pt]{\PastLTLr};
\node (stlr) at (-3,-5) [inner sep=8pt]{\PastSTLr};
\draw [->, thick] (ltl) -- (mtl) node[midway, left] {\textxt{a}{b}};
\draw [->, thick] (mtl) -- (stl) node[midway, left] {$\geq$};
\draw [->, thick] (foltl) -- (fomtl) node[midway, right] {\textxt{a}{b}};
\draw [->, thick] (fomtl) -- (fostl) node[midway, right] {$\geq$};
\draw [->, thick] (ltlr) -- (stlr) node[midway, left] {\textxt{a}{b}};
\draw [->, thick] (ltl) -- (foltl) node[midway, above right] {$\forall\exists$};
\draw [->, thick] (mtl) -- (fomtl) node[midway, above right] {$\forall\exists$};
\draw [->, thick] (stl) -- (fostl) node[midway, above right] {$\forall\exists$};
\draw [->, thick] (stl) -- (stlr) node[midway, above left] {$\rho$};

\draw [->, thick] (ltl) to[out=180,in=90] node [above left]{$\rho$, $\geq$} (ltlr);
\end{tikzpicture}
\end{center}
\caption{Hierarchy of supported temporal logic formalisms in the Reelay framework}
\label{fig:formalisms}
\end{figure}

This paper presents our online monitoring framework, Reelay\footnote{https://github.com/doganulus/reelay}, which unifies temporal logic monitoring under a single computational model and interface.
The Reelay framework provides a comprehensive solution for monitoring \LTL, \MTL, \STL, as well as their robustness and first-order extensions over discrete and dense-time behaviors.
We illustrate these formalisms and the relations between them in Figure~\ref{fig:formalisms}.
Labeled arrows indicate the direction of syntactic or semantic extensions, representing predicates ($\geq$), timing constraints (\textxt{a}{b}), robustness semantics ($\rho$), and quantification over categorical variables ($\forall\exists$).
For all supported formalisms, the Reelay framework strictly adheres to past temporal logic operators to ensure a strictly causal analysis, enabling real-time monitoring and immediate feedback.

Reelay’s unified specification language supports syntactic elements from the supported logical formalisms, enabling the concise expression of complex temporal properties through atomic predicates, temporal operators, timed constraints, categorical variables, and quantifiers.
A defining characteristic of the framework is its versatile runtime evaluation engine, which natively supports both discrete-time and dense-time monitoring.
As described in~\cite{ulus2019online}, this capability allows a single logical specification to be evaluated consistently across diverse execution contexts, ranging from high-frequency real-time message streams to retrospective analysis of logged historical data.
As a result, users can seamlessly switch between time models and data formats without modifying their specifications.

Table \ref{tab:tools-comparison} compares Reelay against modern temporal logic monitoring tools, highlighting a landscape currently fragmented by narrowly targeted temporal logics.
Tools such as DejaVu~\cite{dejavu} and MonPoly~\cite{basin2015monitoring} facilitate first-order temporal monitoring, enabling the specification of properties characterized by data quantification and complex relational structures. DejaVu, however, only supports discrete-time semantics, which may fail to capture the continuous-time dynamics that are critical in cyber-physical systems. While MonPoly supports dense-time, it relies on sample-based evaluation models that inherently couple property evaluation to the underlying sampling scheme. Such coupling may yield inconsistent results in low-frequency asynchronous settings, and may mask violations occurring between sampling instants~\cite{basin2018algorithms}. Furthermore, MonPoly restricts first-order quantification to finite domains, constraining its applicability in data-intensive scenarios involving unbounded domains.
On the other hand, RTAMT~\cite{rtamt} targets dense-time monitoring of continuous signals using \STL, with a primary emphasis on numerical data and robustness semantics. Finally, these tools also differ substantially in their specification languages and input syntaxes, further contributing to the lack of a unified monitoring framework.

\begin{table}[bt]
\caption{Comparison of Reelay with existing monitoring tools across supported formalisms and time models.}
\begin{threeparttable}[]
\begin{tabularx}{\columnwidth}{lccc}
\toprule
Logical &  Discrete & \multicolumn{2}{c}{Dense Time}\\
Formalism & Time & Sample-based & State-based \\
\midrule
\PastLTL    & \ding{51}\tnote{d,p,r} & --- & --- \\
\PastLTLr   & \ding{51}\tnote{r} & --- & --- \\
\PastFOLTL & \ding{51}\tnote{d,p} & --- & --- \\
\PastMTL    & \ding{51}\tnote{d,p,r} &\ding{55}\tnote{p} & \ding{51}\tnote{a,r}\\
\PastFOMTL & \ding{51}\tnote{d,p} &\ding{55}\tnote{p}& \ding{51}\\
\PastSTL    & \ding{51}\tnote{r} &---& \ding{51}\tnote{a,r}\\
\PastSTLr   & \ding{51}\tnote{r} &---& \ding{51}\tnote{r}\\
\PastFOSTL & \ding{51} &---& \ding{51}\\
\bottomrule
\end{tabularx}
\begin{tablenotes}
\item [\ding{51}] Supported in Reelay
\item [\ding{55}] Not supported in Reelay
\item [a] Supported in AMT2~\cite{amt2}
\item [d] Supported in DejaVu~\cite{dejavu}
\item [p] Supported in MonPoly~\cite{basin2015monitoring}
\item [r] Supported in RTAMT~\cite{rtamt}
\end{tablenotes}
\end{threeparttable}
\label{tab:tools-comparison}
\end{table}

The remainder of this paper is organized as follows.
Section~\ref{sec:behaviors} establishes our formal model for temporal behaviors and discusses the rationale behind our practical design choices.
Section~\ref{sec:specification} provides a detailed exposition of the syntax and semantics of the Reelay specification language.
Implementation details and the underlying computational architecture are discussed in Section~\ref{sec:implementation}.
In Section~\ref{sec:example} and~\ref{sec:performance}, we present an illustrative case study and a comprehensive tool performance evaluation.
Finally, Section~\ref{sec:conclusion} concludes the paper with a summary of our findings and outlines potential directions for the future work.

\section{Temporal Behaviors}
\label{sec:behaviors}
This section introduces the temporal behavior model used in the Reelay framework and describes how system executions are represented for online monitoring.

A temporal behavior is a sequence of data messages collected from a real-time system ordered by timestamps.
Reelay assumes that temporal behaviors satisfy a strict monotonicity condition, requiring timestamps to increase strictly between successive messages.
The framework is independent of any particular message format and may support multiple alternatives including comma-separated values (CSV), multi-line JSON documents, and length-delimited binary files, we consistently use newline-delimited JSON throughout this paper.
This choice is motivated by its readability in illustrative examples and its straightforward serialization in practical implementations.

\begin{figure}[b]
\begin{subfigure}[b]{\columnwidth}
\centering
\begin{minted}[
breaklines,
frame=single,
framesep=2mm,
tabsize=2,
xleftmargin=0pt,
samepage=true,
baselinestretch=0.5,
]{js}
{"p1": false, "nd": 1.23, "enm1": "A"} // time: 0
{"p1": true,  "nd": 0.01, "enm1": "A"} // time: 1
{"p1": true,  "nd": 9.12, "enm1": "B"} // time: 2
{"p1": true,  "nd": 9.12, "enm1": "B"} // time: 3
{"p1": false, "nd": 9.18, "enm1": "C"} // time: 4
\end{minted}
\vspace*{-5mm}
\caption{}
\label{fig:discrete-time-behavior-example-1}
\end{subfigure}
\begin{subfigure}[b]{\columnwidth}
\begin{minted}[
breaklines,
frame=single,
framesep=2mm,
tabsize=2,
xleftmargin=0pt,
samepage=true,
baselinestretch=0.5,
]{js}
{"p1": false, "nd": 1.23, "enm1": "A"} // time: 0
{"p1": true,  "nd": 0.01,            } // time: 1
{             "nd": 9.12, "enm1": "B"} // time: 2
{                                    } // time: 3
{"p1": false, "nd": 9.18, "enm1": "C"} // time: 4
\end{minted}
\vspace*{-5mm}
\caption{}
\label{fig:discrete-time-behavior-example-2}
\end{subfigure}
\caption{Example discrete-time temporal behaviors showing (a) full-state and (b) delta-encoded  message representations.}
\label{fig:discrete-time-behavior-example}
\end{figure}

Reelay provides a unified framework that supports both discrete-time and dense-time temporal models.
The discrete-time model assumes a fixed, regular progression of time between successive messages.
In this model, message timestamps are implicit and determined by their position in the sequence, and timing constraints are interpreted with respect to discrete step indices rather than physical time.
We illustrate a discrete-time behavior example in Figure~\ref{fig:discrete-time-behavior-example-1}, defined using three key-value pairs where the key \texttt{p1} is a Boolean variable, \texttt{nd} is a numerical variable, and \texttt{enm1} is a categorical data (string) variable.
Those keys may be later referred to in Reelay expressions, as explained in the following section.
Reelay currently supports all scalar JSON-compatible datatypes in temporal behaviors: Boolean values, numbers, strings, and the \texttt{null} value.

In many real-time systems, state variables exhibit \emph{slowly changing} dynamics, particularly when sampled at high frequencies.
Transmitting the full state in every message under these conditions often results in redundant data and inefficient bandwidth utilization.
To mitigate this, Reelay supports \emph{delta-encoded} temporal behaviors, where messages only transmit fields that have changed since the previous time step.
Reelay treats message data as \emph{persistent}: any key omitted from a message is assumed to retain its last known value until it is explicitly updated or set to \texttt{null}.
Figure~\ref{fig:discrete-time-behavior-example-2} illustrates such a delta-encoded discrete-time behavior, equivalent to the original behavior in Figure~\ref{fig:discrete-time-behavior-example-1}.
In this case, an empty message is sent if no data change occurs, only denoting a time passage on one discrete time unit.

\begin{figure}[t]
\centering
\begin{subfigure}[b]{\columnwidth}
\begin{minted}[
breaklines,
frame=single,
framesep=2mm,
tabsize=2,
xleftmargin=0pt,
samepage=true,
]{js}
{"time": 0,  "p1": false, "nd": 1.23, "enm1": "A"}
{"time": 4,  "p1": true,  "nd": 0.01             }
{"time": 7,               "nd": 9.12, "enm1": "B"}
{"time": 9,  "p1": false, "nd": 9.18, "enm1": "C"}
\end{minted}
\vspace*{-5mm}
\caption{}
\label{fig:dense-time-behavior-example-1}
\end{subfigure}
\begin{subfigure}[b]{\columnwidth}
\begin{minted}[
breaklines,
frame=single,
framesep=2mm,
tabsize=2,
xleftmargin=0pt,
samepage=true,
]{js}
{"p1": false, "nd": 1.23, "enm1": "A"} // time: 0
{}                                     // time: 1
{}                                     // time: 2
{}                                     // time: 3
{"p1": true,  "nd": 0.01}              // time: 4
{}                                     // time: 5
{}                                     // time: 6
{             "nd": 9.12, "enm1": "B"} // time: 7
{}                                     // time: 8
{"p1": false, "nd": 9.18, "enm1": "C"} // time: 9
\end{minted}
\vspace*{-5mm}
\caption{}
\label{fig:dense-time-behavior-example-2}
\end{subfigure}
\caption{(a) Example dense-time temporal behavior and (b) its equivalent discrete-time representation.}
\label{fig:dense-time-behavior-example}
\end{figure}

Conversely, dense-time behaviors support arbitrary, non-uniform time intervals between successive messages.
Unlike the discrete-time model, where timing is derived from sequence position, dense-time messages must include an explicit timestamp.
By default, Reelay expects this value in a top-level \texttt{time} field, though the specific field name can be tailored via Reelay’s configuration.
%
%
%
Figure~\ref{fig:dense-time-behavior-example-1} illustrates a dense-time behavior in newline-delimited JSON format.
For example, this behavior shows the Boolean variable \texttt{p1} is false from time 0 to 4, true from 4 to 9, and false afterward.
Figure~\ref{fig:dense-time-behavior-example-2} presents an equivalent discrete-time behavior for comparison.
In a discrete-time model, representing the same underlying information requires a sequence of messages in which each index corresponds to a uniform clock tick.
Consequently, dense-time models are often better suited for capturing asynchronous, event-driven data and slowly varying signals.

\section{Reelay Expressions}
\label{sec:specification}

Reelay generates runtime monitors dynamically from user-defined specifications, offering a flexible and structured approach to specifying functional temporal properties. These specifications are written in Reelay's expression format, which utilizes a variety of syntactic elements categorized into three types: atomic expressions, Boolean expressions, and temporal expressions. This section outlines the structure of these syntactic elements and provides examples of their usage.

\subsection{Atomic Expressions}

Atomic expressions are the fundamental building blocks of Reelay expressions, representing basic conditions or events that can be monitored over system behaviors. Reelay uses a familiar curly bracket syntax to group a set of constraints over incoming message fields.
For example, we write an atomic expression to check that a Boolean variable \texttt{p1} is true, a numerical variable \texttt{nd} is greater than a constant value of \texttt{9.0}, and a categorical (string) variable \texttt{enm1} equals the string of \texttt{B} as follows:
\begin{center}
\begin{BVerbatim}[baseline=c]
{p1: true, nd > 9.0, enm1: "B"}
\end{BVerbatim}
\end{center}
Atomic expressions will evaluate to true only if all conditions are met simultaneously for a time point. For instance, given the discrete-time behavior illustrated in Figure~\ref{fig:discrete-time-behavior-example}, the expression above evaluates to \texttt{false} at time points 0, 1, and 4, while evaluating to \texttt{true} at time points 2 and 3.

Moreover, Reelay allows users to define basic predicates using Boolean checks,  numerical comparisons, and string equality checks. For more advanced tasks, the user can incorporate their custom predicates by using \texttt{\small\$\{func1\}} syntax where \texttt{\small func1} is the registered name of a Boolean-valued function over the input message. This extensibility is natively supported in both the C++ and Python libraries, providing a flexible architecture for incorporating complex domain-specific logic that exceeds the scope of the core specification language.

\subsection{Boolean Operators}

Reelay provides a standard set of Boolean connectives to compose complex logical properties. These operators follow conventional precedence rules and are evaluated pointwise at each time point or over intervals, depending on the underlying time model.
Each operator is presented below with its description, a representative example, and its semantic interpretation.

\paragraph{Negation.}
The negation operation is specified using the unary prefix connectives \texttt{not} or \texttt{!}. For example, the expression
\begin{center}
\begin{BVerbatim}[baseline=c]
not {key1: value1, key2: value2}
\end{BVerbatim}
\end{center}
This expression evaluates to \texttt{true} if and only if the nested atomic expression is \texttt{false}. Under standard semantics, this requires that the conjunction of the inner constraints is not satisfied at the time of evaluation.

\paragraph{Conjunction.}
The conjunction operation is specified using the binary connectives \texttt{and} or \texttt{\&\&}. For example, the expression
\begin{center}
\begin{BVerbatim}[baseline=c]
{key1: value1} and {key2: value2}
\end{BVerbatim}
\end{center}
holds only if both operands are satisfied simultaneously. Notably, the curly-bracket syntax for atomic expressions is a shorthand notation for conjunctions over individual message fields. Consequently, the example above is logically equivalent to the single atomic expression \texttt{\string{key1: value1, key2: value2\string}}.

\paragraph{Disjunction.}

The disjunction operation is specified using the binary connectives \texttt{or} or \texttt{||}. For example, the expression
\begin{center}
\begin{BVerbatim}[baseline=c]
{key1: value1} or {key2: value2}
\end{BVerbatim}
\end{center}
evaluates to \texttt{true} if at least one of the atomic expressions is satisfied at the current time point.

\paragraph{Implication.}
The logical implication operation is specified using the binary connectives \texttt{implies} or \texttt{->}. For example, the expression
\begin{center}
\begin{BVerbatim}[baseline=c]
{key1: value1} implies {key2: value2}
\end{BVerbatim}
\end{center}
follows standard Boolean semantics and is semantically equivalent to \texttt{not\string{key1: value1\string} or \string{key2: value2\string}}.

\subsection{Temporal Operators}

Reelay provides a standard set of past temporal logic connectives to compose complex logical properties over the history of a behavior.
These operations enable the precise specification of requirements that depend on previously observed states or events.
Each operator is presented below with an informal description, a representative example, and an explanation of its semantic interpretation.
Notably, Reelay adopts slightly different semantics for these operators under discrete-time and dense-time models.
Formal definitions of the past temporal operators are provided in~\cite{ulus2019online}.

\paragraph{Previously.}
The previously operation is specified using the unary prefix connectives \texttt{pre} or \texttt{Y}. For example, the expression
\begin{center}
\begin{BVerbatim}[baseline=c]
pre {key1: value1, key2: value2}
\end{BVerbatim}
\end{center}
evaluates to \texttt{true} at the current time point if the nested expression was satisfied at the immediate preceding time point. Note that the \texttt{pre} operation is uniquely meaningful in discrete-time settings, as it refers to the specific previous message in the sequence.

\paragraph{Once.}
The once (or sometime in the past) operation is specified using the unary prefix connectives \texttt{once} or \texttt{P}. For example, the expression
\begin{center}
\begin{BVerbatim}[baseline=c]
once {key1: value1, key2: value2}
\end{BVerbatim}
\end{center}
evaluates to \texttt{true} if the nested expression has held at least once at some point in the past, including the current time point.

\paragraph{Historically.}
The historically (or always in the past) operation is specified using the unary prefix connectives \texttt{always} or \texttt{H}. For example, the expression
\begin{center}
\begin{BVerbatim}[baseline=c]
always {key1: value1, key2: value2}
\end{BVerbatim}
\end{center}
requires that the nested expression has held at every time point from the beginning of the behavior up to the current moment.

\paragraph{Since.}
The since operation is a binary temporal operation specified using the connectives \texttt{since} or \texttt{S}. For example, the expression
\begin{center}
\begin{BVerbatim}[baseline=c]
{key1: value1} since {key2: value2}
\end{BVerbatim}
\end{center}
evaluates to \texttt{true} if the second operand held at some time point in the past, and the first operand has held continuously from that point until the current time.

\subsection{Variables and Quantifiers}
Reelay fully supports first-order temporal logic over unbounded categorical data domains under both discrete-time and dense-time models, with the current exception of robustness semantics.
This capability enables the specification of properties that quantify over dynamically observed data values within temporal behaviors, where the data domain is incrementally discovered and updated as new data arrives.
Reelay handles data values symbolically, enabling efficient reasoning over dynamically evolving domains. Formal definitions of the first-order past temporal logic, along with the corresponding algorithmic details implemented, are provided in~\cite{havelund2020first}.

\paragraph{Reference Variables.}
Reelay enables the specification of properties involving data values unknown at compile time through the use of \textit{reference variables}.
A reference variable acts as a symbolic placeholder that binds to a specific data value from a message field, allowing that value to be stored and subsequently compared for equality within the same or across different time points.
The syntax \texttt{*var} is used within an atomic expression to declare or reference a bound value, where \texttt{var} is the variable name.
Currently, this support is available exclusively for categorical (string) variables.

\paragraph{Quantification Syntax.}
The universal (\texttt{forall}) and existential (\texttt{exists}) quantifiers enable the specification of properties that range over the dynamically discovered, unbounded domain of categorical values encountered throughout a behavior's history.
Unlike propositional temporal logic, these quantifiers allow Reelay to reason about an arbitrary number of distinct data values, such as identifiers or entity names, that appear within the data stream.
The syntax for defining these quantifiers and their associated variables is as follows:
\begin{center}
\begin{BVerbatim}[baseline=c]
exists[v1, v2, ...]. RYEXPR(v1, v2, ...)
\end{BVerbatim}
\end{center}
\begin{center}
\begin{BVerbatim}[baseline=c]
forall[v1, v2, ...]. RYEXPR(v1, v2, ...)
\end{BVerbatim}
\end{center}
where \texttt{RYEXPR(v1, v2, ...)} is an arbitrary Reelay expression containing the variable declarations.
For instance, the following specification uses the existential quantifier to verify the equality of two message fields whose concrete values are not known \textit{a priori}:
\begin{center}
\begin{BVerbatim}[baseline=c]
exists[var]. {key1: *var, key2: *var}
\end{BVerbatim}
\end{center}
This expression evaluates to \texttt{true} for a data message such as \texttt{\string{"key1": "hello", "key2": "hello"\string}}, as the constant \texttt{"hello"} provides a valid assignment for the variable \texttt{var} that satisfies both atomic constraints.
Conversely, the expression evaluates to \texttt{false} for the message \texttt{\string{"key1": "hello", "key2": "world"\string}}, since the field values cannot be unified under a single assignment for the variable \texttt{var}.

\subsection{Operator Precedence}
All binary temporal operators in Reelay are left associative.
When temporal and Boolean operators are combined without explicit parentheses, expressions are evaluated according to a fixed precedence hierarchy.
Unary operators such as \texttt{not}, \texttt{pre}, \texttt{once}, and \texttt{always} have the highest binding strength, followed by conjunction (\texttt{and}), disjunction (\texttt{or}), and binary temporal operators such as \texttt{since}.

\section{Software Design and Implementation}
\label{sec:implementation}

Reelay is designed with a strong emphasis on performance, flexibility, and modularity, providing a robust library for real-time, system-level verification. Its core is implemented in \texttt{C++} to achieve high performance, enabling both online monitoring of live message streams and efficient offline analysis of recorded execution logs. This paper focuses on the framework’s tooling and implementation details, building upon the formal foundations and algorithmic analyses presented in~\cite{ulus2019online}.

\subsection{Sequential Networks as Runtime Engine}
The monitoring process begins with specifications defined in the Reelay expression format. Rather than interpreting these formulas directly at runtime, Reelay adopts a compiler-like approach that translates each formula into a \emph{sequential network}, realized as a synchronous computation graph~\cite{ulus2019online}. This graph is constructed by recursively decomposing the specification into its constituent subexpressions, where each node represents a specific logical operator.

During the translation phase, well-known optimization techniques such as common subexpression elimination are applied to detect and merge redundant nodes. By identifying identical subformulas and mapping them to shared nodes, the framework ensures that repetitive operations are evaluated only once per time step, reducing both memory footprint and computational overhead. Consequently, the computation graph serves not merely as a structural representation of the specification, but as an optimized, high-throughput execution engine.

\subsection{Algebraic Data Structures}

The Reelay framework employs a specialized collection of algebraic data structures to implement the semantics of its supported logical formalisms while sustaining high-performance runtime evaluation. In this architecture, \LTL operators are evaluated over the Boolean domain~\cite{havelund2004monitoring}, whereas timed \MTL operators leverage interval-set representations over the time domain to efficiently handle timing constraints~\cite{ulus2019online}. Robustness semantics are defined over min-max algebra over numeric fields~\cite{donze2010robust}, while first-order reasoning over unbounded categorical domains is realized using Binary Decision Diagrams (BDDs)~\cite{havelund2020first}. Reelay further integrates temporal and quantitative reasoning by composing these individual domains through product algebras, allowing heterogeneous semantics to be evaluated uniformly within a runtime.

\subsection{Compile-Time Specialization for Message Formats}
A distinctive aspect of Reelay’s extensibility is its ability to adapt to different message input formats at compile time. Through the use of C++ template specialization, the library allows developers to define type-specific handling for a wide range of data representations, including complex user-defined structures, without modifying the core monitoring logic.

Once a temporal logic specification has been compiled into a computation graph, this approach enables the compiler to generate a monitor that is statically optimized for the concrete data structures being processed. As a result, input handling and data access are resolved at compile time rather than through runtime dispatch or interpretation. This design prevents message field access from becoming a performance bottleneck and embeds efficient, type-specific processing paths directly into the nodes of the computation graph, thereby maximizing throughput and minimizing runtime overhead.

\subsection{Header-Only C++ Architecture}
Reelay is implemented as a header-only \texttt{C++17} library.
This design choice enables aggressive compile-time optimizations, including cross-translation-unit inlining and specialization, thereby minimizing runtime overhead. Leveraging modern \texttt{C++} abstractions, the library maintains a small memory footprint and a predictable, low-latency execution profile, making it well suited for the stringent resource and timing constraints of real-time system verification.

\subsection{Python Integration and Wrapper Layer}
Reelay bridges the gap between high-level usability and low-level performance through a standalone Python library.
This interface is built using \texttt{pybind11}, which enables a seamless, high-efficiency connection between Python's abstractions and the underlying C++ core.

Users can define temporal properties in a declarative style and manage data streams using familiar Python tools.
While the API feels natively Pythonic, the construction of the computation graph and the incremental state updates are handled entirely by the C++ engine.
This design allows for rapid prototyping and testing without sacrificing the execution speed required to process massive datasets or high frequency telemetry.

\section{Practical Usage}
\label{sec:example}

This section demonstrates the practical application of Reelay by checking the functional requirements of a Door Open Warning (DOW) feature for an autonomous robotic home assistant.
The feature is designed to alert users if an entry door remains open for an extended duration while suppressing redundant notifications to ensure a non-intrusive user experience.
We present the formalization of these requirements and their implementation in both C++ and Python.

\subsection{Designing Door Open Warning Feature}

The development process begins by decomposing high-level functional objectives into precise, verifiable system requirements.
This step is essential for bridging the gap between informal, human-readable feature descriptions and the rigor demanded by formal runtime monitoring frameworks.
By explicitly capturing timing constraints and event dependencies at the requirement level, the resulting specifications become both executable and directly amenable to formal reasoning.

For the Door Open Warning (DOW) feature, we define two core requirements that characterize the system's functionality:
\begin{description}
\item[\texttt{DOW-REQ-01}] The system shall issue a warning if the entry door remains open for at least 5 minutes.
\item[\texttt{DOW-REQ-02}] The door shall be closed before the system issues another warning. This should prevent the system from issuing multiple warnings while the door remains open.
\end{description}

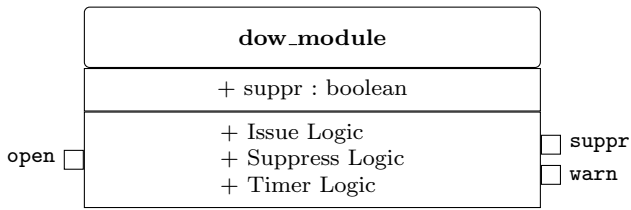
\begin{figure}[b]
\centering
\begin{tikzpicture}[
    uml/.style={
        draw,
        rectangle,
        rounded corners=2pt,
        align=left,
        minimum width=6cm
    },
    title/.style={
        draw,
        rectangle,
        rounded corners=2pt,
        minimum width=6cm,
        minimum height=0.8cm,
        fill=white
    },
    compartment/.style={
        draw,
        rectangle,
        minimum width=6cm,
        fill=white
    },
    port/.style={
        draw,
        rectangle,
        minimum width=0.25cm,
        minimum height=0.25cm
    }
]

\node[title] (title) {\centering \textbf{dow\_module}};

\node[compartment, below=0pt of title] (attr) {
\begin{tabular}{l}
$+$ suppr : boolean
\end{tabular}
};

\node[compartment, below=0pt of attr] (logic) {
\begin{tabular}{l}
$+$ Issue Logic \\
$+$ Suppress Logic \\
$+$ Timer Logic
\end{tabular}
};

\node[port, left=0mm of logic.west] (inport) {};
\node[port, right=0mm of logic.east, yshift=2mm] (outsuppr) {};
\node[port, right=0mm of logic.east, yshift=-2mm] (outwarn) {};

\node[left=0mm of inport] {\texttt{open}};
\node[right=0mm of outwarn] {\texttt{warn}};
\node[right=0mm of outsuppr] {\texttt{suppr}};

\end{tikzpicture}
\caption{Block diagram of the Door Open Warning (DOW) module.}
\label{fig:dow_diagram}
\end{figure}

\noindent Figure~\ref{fig:dow_diagram} illustrates the logical architecture of the Door Open Warning (DOW) system.
The \texttt{dow\_module} evaluates the system state by observing the \texttt{open} signal, which represents the physical door status, together with the \texttt{suppr} flag, which captures alarm suppression.
Internally, the module combines timer logic to measure the duration of door-open events, issue logic to detect violations of the specified temporal threshold, and suppression logic to control alarm recurrence.
System executions are modeled as sequences of observable states over the module’s input and output interfaces.
Figure~\ref{fig:dow_observations} shows an example discrete-time execution trace derived from these interfaces, in which the system samples data at fixed one-minute intervals.
The trace illustrates how a sustained door-open condition leads to the issuance of a warning, followed by activation of suppression to inhibit further alerts while the condition persists.

\begin{figure}[tb]
\begin{minted}[frame=single, fontsize=\footnotesize]{js}
{"open": false, "suppr": false, "warn": false}
{"open": true,  "suppr": false, "warn": false}
{"open": true,  "suppr": false, "warn": false}
{"open": true,  "suppr": false, "warn": false}
{"open": true,  "suppr": false, "warn": false}
{"open": true,  "suppr": false, "warn": false}
{"open": true,  "suppr": false, "warn": true }
{"open": true,  "suppr": true,  "warn": false}
{"open": true,  "suppr": true,  "warn": false}
\end{minted}
\caption{Example discrete-time behavior for the Door Open Warning (DOW) module.}
\label{fig:dow_observations}
\end{figure}

\subsection{Formalizing Door Open Warning Feature}
In this section, we formalize the DOW requirements using the Reelay expression format.

The first requirement, \texttt{DOW-REQ-01}, states that the system shall issue a warning if the door remains open for at least five minutes.
This requirement must be interpreted in conjunction with \texttt{DOW-REQ-02}, which constrains warning issue logic when suppression is active.
While such dependencies are often implicit in informal specifications, they must be stated explicitly in automated verification.
The first part of \texttt{DOW-REQ-01} is expressed in the Reelay expression format as follows:
\begin{center}
\begin{BVerbatim}[baseline=c]
(H[0:5]{open} and not{suppr}) -> {warn}
\end{BVerbatim}
\end{center}
This formula enforces that a warning must be raised whenever the door has been continuously open for the preceding five minutes and warning suppression is inactive. Violating this property means a missing warning.

In natural language, conditional statements often implicitly convey a bidirectional interpretation. To reflect this intent formally, \texttt{DOW-REQ-01} is complemented with additional constraints that rule out false positives by requiring that any issued warning be justified by the corresponding door-open condition and the absence of suppression:
\begin{center}
\begin{BVerbatim}[baseline=c]
{warn} -> H[0:5]{open}
\end{BVerbatim}
\end{center}
and the current state of the suppression signal:
\begin{center}
\begin{BVerbatim}[baseline=c]
{warn} -> not{suppr}
\end{BVerbatim}
\end{center}
Together, these properties align the system's output with its triggers, establishing the logical equivalence necessary for precise verification.

The second requirement, \texttt{DOW-REQ-02}, specifies a timing constraint intended to prevent repeated warnings for a single continuous door-open event. This requirement involves more expressive temporal reasoning and makes use of the \texttt{pre} and \texttt{since} operators:
\begin{center}
\begin{BVerbatim}[baseline=c]
{warn} -> not(pre({open} since {warn}))
\end{BVerbatim}
\end{center}
This specification states that whenever a warning is issued, the door must not have remained continuously open since the previous warning. In other words, the door must be closed between two warning events. A violation of this property indicates a false positive and directly captures the intended semantics of \texttt{DOW-REQ-02}.

\subsection{Reelay Monitors in C++}

In this section, we demonstrate how to instantiate and execute Reelay monitors in C++, following the formalization of the Door Open Warning (DOW) requirements.
These monitors process system behaviors in real-time or reading from logs and evaluate them against the temporal logic specifications to detect violations.

The library utilizes \texttt{options} and \texttt{make\_monitor} constructs to define the monitor's operational environment. To configure a discrete-time monitor, we first define an \texttt{options} object to specify the underlying time representation, data types, and output behavior:
\begin{minted}[frame=lines, framesep=2mm, fontsize=\footnotesize]{cpp}
auto opts = reelay::discrete_timed<intmax_t>
    ::monitor<reelay::json, reelay::json>::options()
        .disable_condensing();
\end{minted}
With the configuration established, the monitor is instantiated by passing the RYE pattern and the \texttt{opts} object to the \texttt{make\_monitor} factory function. Note that a single \texttt{opts} instance can be reused to create multiple monitors sharing the same configuration profile.
\begin{minted}[frame=lines, framesep=2mm, fontsize=\footnotesize]{cpp}
std::string s = "(H[0:5]{open} && !{suppr})->{warn}";
auto m = reelay::make_monitor(s, opts);
\end{minted}
The monitor’s internal state is advanced via the \texttt{update} method, which processes input incrementally. A common implementation pattern involves iterating through a newline delimited logfile, parsing each entry, and evaluating the verdict as follows:

\begin{minted}[frame=lines, framesep=2mm, fontsize=\footnotesize]{cpp}
for (std::string line; std::getline(logfile, line);){
    reelay::json msg = reelay::json::parse(line);
    auto result = m.update(msg);

    if (result["value"] == false) {
        std::cout << "Error at " << m.now()
                  << " : Violation!"
                  << std::endl;
    }
}
\end{minted}
As the monitor traverses the input stream, it assesses the temporal logic relative to the current system time. Because these safety requirements are expected to hold globally, any \texttt{false} verdict indicates a formal violation of the specification, printing an immediate report. Other properties, such as those defining warning suppression or complex temporal dependencies, are instantiated and executed similarly.

\subsection{Reelay Monitors in Python}

In this section, we demonstrate how to instantiate and execute Reelay monitors in Python. The Python API mirrors the C++ core by providing two constructs for discrete and dense-time monitors. Unlike the C++ API, which requires explicit template instantiation for data types, the Python monitor is designed to work natively with standard Python dictionaries. Configuration is handled conveniently through keyword arguments.

To build a monitor for \texttt{DOW-REQ-01}, we pass the Reelay expression directly to the constructor as follows:

\begin{minted}[frame=lines, framesep=2mm, fontsize=\footnotesize]{python}
s = r"(H[0:5]{open} and not {suppr}) -> {warn}"
m = reelay.discrete_timed_monitor(s, condense=False)
\end{minted}

The monitor’s internal state is updated via the \texttt{update} method, which accepts a dictionary representing the signal values at the current time step. If the system behavior is stored as a list of dictionaries, we can iterate through the trace as follows:

\begin{minted}[frame=lines, framesep=2mm, fontsize=\footnotesize]{python}
for msg in system_behavior:
    result = m.update(msg)
    if result["value"] is False:
        print(f"Error at {m.now()}: Violation!")
\end{minted}
The \texttt{result} variable is a dictionary containing the truth value of the specification. Because the safety requirements are expected to hold globally, any \texttt{False} verdict indicates a formal violation, triggering an immediate report.

\section{Performance Evaluation}
\label{sec:performance}

This section demonstrates the performance and scalability of Reelay monitors and compares them against existing runtime monitoring tools.
The primary objective is to quantify the computational overhead of various forms of temporal logic monitoring, specifically focusing on metric timing constraints, robustness semantics, and first-order quantification.


The performance evaluation uses the Timescales \cite{ulus2019timescales} and DejaVu \cite{dejavu-benchmark} benchmark suites.
Together, these benchmarks provide a diverse collection of execution logs and temporal logic specifications that employ different fragments of past temporal logic, including \MTL, \STL, and first-order extensions.
We evaluate tool performance based on per-message processing time, calculated by dividing the total execution time required for log analysis by the number of messages processed.
For consistency, every log file used in our experiments contains approximately one million messages.
All benchmarks are conducted on a 3.80GHz Intel Xeon W-2235 CPU with 32GB memory running Linux.

\subsection{Base Performance}

We first evaluate the performance of \PastMTL monitoring to establish a baseline for other monitoring applications. Rather than operating as a monolithic tool, the Reelay framework enables the development of various temporal logic monitoring applications tailored to specific deployment contexts and message formats.
By comparing these variants, we can isolate the performance trade-offs associated with message representation from the intrinsic efficiency of the underlying temporal logic engine.

In particular, we characterize the performance of three different Reelay-based command-line interfaces: (1) \texttt{ryjson}, a C++ application utilizing the core Reelay library and the \texttt{simdjson} library\footnote{\texttt{https://github.com/simdjson/simdjson}} for high performance JSON parsing~\cite{simdjson}; (2) \texttt{ryjson-py}, a Python application employing the Reelay Python bindings and builtin \texttt{json} module; and (3) \texttt{rybinx}, a C++ application operating directly on binary serializations of native message structs.
While monitoring over multi-line JSON documents offers cross-language portability and message format flexibility, it introduces significant overhead due to runtime string parsing and dynamic type resolution.
In contrast, the native C++ implementation (\texttt{rybinx}) eliminates these deserialization costs entirely, thereby demonstrating the framework’s effective upper bound on achievable performance.

Table~\ref{tab:serialization_perf} quantifies these performance trade-offs over selected Timescales properties.
Our experiments demonstrate that the primary application, \texttt{ryjson}, is highly performant, processing JSON messages in the submicrosecond range with variations depending on property complexity.
The C++ implementation, \texttt{rybinx}, achieves a significant performance boost by utilizing native C++ structs, often reducing processing time by an additional 50--60\% compared to \texttt{ryjson}.
However, these structs are hard-coded and require schema-specific tailoring at compile time.
Conversely, \texttt{ryjson-py} exhibits the highest overhead per message; this is attributed to the performance gap between implementation languages, as the Python bindings must manage native Python objects.
Importantly, the Reelay engine demonstrates excellent scalability. When evaluating properties with timing constraints spanning two orders of magnitude (10 to 1000), the processing latency remains nearly constant.
This suggests that the engine's overhead is independent of the magnitude of timing bounds in the properties being monitored.
As detailed in \cite{ulus2019online}, this constant-time performance is achieved by representing time intervals symbolically.

\begin{table}[tb]
\centering
\caption{Average per-message processing time across three Reelay applications.}
\label{tab:serialization_perf}
\begin{tabularx}{\columnwidth}{>{\ttfamily\footnotesize}X rrrr}
\toprule
\ttfamily Property & \ttfamily ryjson-py & \ttfamily ryjson & \ttfamily rybinx \\
\midrule
AbsentAQ10     & 2468 ns & \bfseries 197 ns & 87 ns\\
AbsentAQ100    & 2453 ns & \bfseries 186 ns & 76 ns\\
AbsentAQ1000   & 2497 ns & \bfseries 182 ns & 75 ns\\
\midrule
AlwaysBR10     & 2221 ns & \bfseries 212 ns & 118 ns \\
AlwaysBR100    & 2252 ns & \bfseries 211 ns & 118 ns \\
AlwaysBR1000   & 2182 ns & \bfseries 211 ns & 119 ns \\
\midrule
RecurBQR10     & 3124 ns & \bfseries 350 ns & 139 ns \\
RecurBQR100    & 3099 ns & \bfseries 319 ns & 110 ns \\
RecurBQR1000   & 3154 ns & \bfseries 312 ns & 109 ns \\
\midrule
RespondBQR10   & 3717 ns & \bfseries 480 ns & 206 ns \\
RespondBQR100  & 3600 ns & \bfseries 443 ns & 173 ns \\
RespondBQR1000 & 3613 ns & \bfseries 433 ns & 162 ns \\
\bottomrule
\end{tabularx}
\end{table}

\subsection{Performance of Robustness Monitoring}

Secondly, we evaluate the performance of \PastSTL monitoring under robustness semantics, comparing Reelay with the \texttt{rtamt} framework, a specialized Python-based STL monitoring library~\cite{rtamt}. For this evaluation, we adapted the Timescales benchmark properties to a quantitative context. Boolean signals were converted to real-valued signals by mapping \texttt{true} and \texttt{false} constants to \texttt{1.1} and \texttt{-1.1}, respectively. Furthermore, atomic propositions within the Timescales properties were replaced with corresponding predicates of the form \texttt{\string{p > 0\string}}.

Table \ref{tab:robustness_perf} demonstrates a significant performance gap between the two frameworks.
For this comparison, we implemented a Python application, \texttt{rtamt-app}, which utilizes the \texttt{rtamt} Python interface and mirrors the implementation of \texttt{ryjson-py} to process multi-line JSON logs.
The C++ implementation, \texttt{ryjson}, is included to highlight the efficiency of the underlying engine without the overhead of the Python interpreter.

While \texttt{rtamt} provides a flexible testing environment for rapid prototyping, its timed monitoring approach relies on the explicit enumeration of time intervals.
This dependency makes it highly sensitive to the magnitude of timing constraints; as formula bounds increase, \texttt{rtamt-app} experiences substantial latency or fails to complete execution (marked as \textsc{dnf}).
In contrast, the Reelay framework provides a similarly accessible rapid prototyping environment through its Python bindings while leveraging high-performance temporal logic monitors.
Notably, Reelay monitors do not experience a slowdown under robustness semantics similar to the Boolean semantics, maintaining consistent performance even as timing constraints scale.

\begin{table}[tb]
\centering
\caption{Average per-message processing time across temporal logic monitors under robustness semantics}
\label{tab:robustness_perf}
\begin{tabularx}{\columnwidth}{>{\ttfamily\footnotesize}X rrr}
\toprule
\ttfamily Property & \ttfamily rtamt-app & \ttfamily ryjson-py & \ttfamily ryjson  \\
\midrule
RoAbsentAQ10     & 20590 ns & 2738 ns & \bfseries 454 ns\\
RoAbsentAQ100    & 31536 ns & 2867 ns & \bfseries 444 ns\\
RoAbsentAQ1000   & 143478 ns & 2818 ns & \bfseries 447 ns\\
\midrule
RoAlwaysBR10     & 15183 ns & 2580 ns & \bfseries 408 ns\\
RoAlwaysBR100    & 26077 ns & 2645 ns & \bfseries 410 ns\\
RoAlwaysBR1000   & 144123 ns & 2516 ns & \bfseries 417 ns\\
\midrule
RoRecurBQR10     & 37396 ns & 3533 ns & \bfseries 605 ns\\
RoRecurBQR100    & 47117 ns & 3512 ns & \bfseries 597 ns\\
RoRecurBQR1000   & 164822 ns & 3430 ns & \bfseries 594 ns\\
\midrule
RoRespondBQR10   & \textsc{dnf} & 5009 ns& \bfseries 1427 ns\\
RoRespondBQR100  & \textsc{dnf} & 4981 ns & \bfseries 1408 ns\\
RoRespondBQR1000 & \textsc{dnf} & 5177 ns & \bfseries 1411 ns\\
\bottomrule
\end{tabularx}
\footnotesize\raggedright
\textsc{dnf}: Did not finish within timeout
\end{table}

\subsection{Performance of First-Order Monitoring}
Finally, we evaluate the performance of first-order temporal logic monitoring by comparing Reelay with the state-of-the-art monitoring tool \texttt{dejavu}~\cite{dejavu}. Our evaluation considers both untimed and timed benchmarks from the \texttt{dejavu} suite over a discrete-time domain. To facilitate a direct comparison, we adapted the original \texttt{dejavu} properties and Comma Separated Value (CSV) traces into equivalent Reelay specifications and multi-line JSON logs, respectively. We assess performance based on the total processing time required for each log file in these experiments.

For untimed monitoring, Reelay and Dejavu use the same underlying technique and algorithm, which employ Binary Decision Diagrams (BDDs) to represent unbounded categorical data domains~\cite{havelund2020first}.
As shown in Table~\ref{tab:first-order-perf}, Reelay demonstrates a consistently better performance across untimed (\PastFOLTL) properties.
This margin may be attributable to implementation level differences: Reelay is implemented in C++ and utilizes the high-performance \texttt{CUDD} library, whereas Dejavu is implemented in Scala and relies on \texttt{JavaBDD}.

On the other hand, the tools diverge more significantly in their handling of first-order timed monitoring (\PastFOMTL).
While Dejavu extends its BDD-based encoding to represent timing constraints~\cite{dejavu-timed}, Reelay generalizes its existing timed monitoring architecture to the first-order setting.
This architectural difference is reflected in their respective scalability as timing constraints increase.
Although Reelay also exhibits sensitivity to the magnitude of timing constraints, the impact is significantly milder than that observed in Dejavu, which fails to complete both benchmarks before the timeout.


\begin{table}[bt]
\centering
\caption{Average per-message processing time across first-order temporal logic monitors}
\label{tab:first-order-perf}
\begin{tabularx}{\columnwidth}{>{\ttfamily\footnotesize}X rrr}
\toprule
\ttfamily Property & \ttfamily dejavu & \ttfamily ryjson \\
\midrule
Access    & 19457 ns& \bfseries 15085 ns\\
Commands & 3793 ns & \bfseries 1016 ns\\
FileOps  & 13430 ns& \bfseries 9572 ns\\
\midrule
LocksBasic    & 11900 ns & \bfseries 4169 ns\\
LocksDataraces & 9249 ns& \bfseries 5900 ns\\
LocksDeadlocks & 36617 ns& \bfseries 3975 ns\\
\midrule
TimedAccess25-50  & 13676 ns& \bfseries 5442 ns\\
TimedAccess25-60  & 99224 ns& \bfseries 6094 ns\\
TimedAccess25-70  & \textsc{dnf} & \bfseries 6527 ns\\ 
\midrule
TimedCommands25-50 & 7927 ns & \bfseries 5434 ns\\
TimedCommands25-60 & 7940 ns & \bfseries 5488 ns\\
TimedCommands25-70 & \textsc{dnf} & \bfseries 5492 ns\\ 
\bottomrule
\end{tabularx}
\footnotesize\raggedright
\textsc{dnf}: Did not finish within timeout
\end{table}

\section{Conclusion}
\label{sec:conclusion}

In this paper, we presented Reelay, a unified framework for the online monitoring of temporal logic specifications. By addressing the fragmentation of existing tools, Reelay provides a single computational model that seamlessly supports a wide array of formalisms, including \PastLTL, \PastMTL, and \PastSTL, alongside their first-order and robustness extensions.

The implementation highlights a successful balance between high-level usability and performance-critical execution. The header-only C++ architecture and computation graph based runtime engine ensure that monitoring overhead remains minimal. Furthermore, the use of template specializations and a standalone Python interface ensures that Reelay is as accessible to data scientists for offline analysis as it is to embedded systems engineers for customized usage in resource constrained environments.

As cyber-physical systems trend toward greater autonomy, the need for expressive and efficient formally specified runtime monitors becomes increasingly critical. Reelay provides the modularity required to meet these demands, offering a versatile foundation for runtime verification in both academic research and industrial applications. Future work will involve extending the framework to support multi-property monitoring architectures and developing decentralized monitoring techniques, specifically targeting the pub/sub architectures prevalent in modern Internet of Things (IoT) and robotics applications.

\bibliographystyle{plain}
\bibliography{references}




\end{document}